# Implementing Zero Trust Architecture to Enhance Security and Resilience in the Pharmaceutical Supply Chain

Saeid Ghasemshirazi [a,*], Ghazaleh Shirvani [a], Marziye Ranjbar Tavakoli [b], Bahar Ghaedi [c], Mohammad Amin Langarizadeh [b]

[a] *Department of Computer Science, Carleton University, Canada, Ottawa.*

[b] *Department of Medicinal Chemistry, Faculty of Pharmacy, Kerman University of Medical Sciences, Kerman, Iran.*

[c] *Student Research Committee, Faculty of Pharmacy, Kerman University of Medical Sciences, Kerman, Iran.*

**Abstract**

The pharmaceutical supply chain faces escalating cybersecurity challenges threatening patient safety and operational continuity. This paper examines the transformative potential of zero trust architecture for enhancing security and resilience within this critical ecosystem. We explore the challenges posed by data breaches, counterfeiting, and disruptions and introduce the principles of continuous verification, least-privilege access, and data-centric security inherent in zero trust. Real-world case studies illustrate successful implementations. Benefits include heightened security, data protection, and adaptable resilience. As recognized by researchers and industrialists, a reliable drug tracing system is crucial for ensuring drug safety throughout the pharmaceutical production process. One of the most pivotal domains within the pharmaceutical industry and its associated supply chains where zero trust can be effectively implemented is in the management of narcotics, high-health-risk drugs, and abusable substances. By embracing zero trust, the pharmaceutical industry fortifies its supply chain against constantly changing cyber threats, ensuring the trustworthiness of critical medical operations.

**Keywords:** Pharmaceutical industry, Drug supply chain, Security, Zero Trust

## Introduction

The pharmaceutical supply chain's efficiency and reliability are paramount, as any disturbance has the potential to have extensive impacts on the well-being and security of the public. However, the pharmaceutical supply chain's increasingly interconnected and digitized nature has exposed it to many cybersecurity threats and vulnerabilities, necessitating innovative approaches to safeguard its integrity and resilience (1,2).

Recently, the idea of "Zero Trust" has gained traction as a promising approach in cybersecurity, challenging the conventional security model based on perimeters. Zero trust is grounded in the fundamental belief that organizations should not trust any user or device solely based on their location within the network. Instead, Continuous verification, strict access controls, and continuous monitoring are required by a zero trust architecture, intending to prevent, detect, and mitigate threats at every stage of interaction (3–5).

This paper addresses the pressing need for improved security and resilience in the pharmaceutical supply chain by advocating for implementing zero trust architecture. By leveraging the principles and strategies inherent in the zero trust approach, the pharmaceutical industry can effectively reduce the risks associated with cyberattacks, data breaches, and supply chain disruptions. The integration of zero trust principles can revolutionize supply chain security, bolstering its capacity to withstand evolving and sophisticated cyber threats (6,7).

In this context, the current paper comprehensively explores the use of zero trust architecture within the pharmaceutical supply chain. Through a critical analysis of the challenges posed by existing security models and an in-depth examination of the principles underpinning zero trust, Our goal is to offer insightful information about the approach's capacity for transformation. Moreover, the paper will present practical strategies and considerations for implementing zero trust within the pharmaceutical supply chain, highlighting real-world case studies and examples to underscore its effectiveness.

As the pharmaceutical industry grapples with the imperatives of digital transformation and heightened cybersecurity risks, the integration of zero trust architecture offers a proactive and adaptable solution. By adopting a zero trust approach, stakeholders within the pharmaceutical supply chain can strengthen their defenses, safeguard critical assets, and guarantee the continuous flow of medications to those who depend on them. This paper thus serves as a foundational guide for industry practitioners, policymakers, and researchers interested in forging a more secure and resilient pharmaceutical supply chain through the principles of zero trust (8,9).

## Background & Related Works

The supply chain management landscape has undergone a paradigm shift in recent decades, with the integration of advanced technologies and digital systems becoming integral to its operations. The pharmaceutical supply chain, characterized by its complexity, global reach, and criticality to public health, is no exception to this transformation. However, alongside the benefits of digitization and globalization, the pharmaceutical supply chain has become increasingly susceptible to a wide variety of cybersecurity threats, including data breaches, counterfeit medications, and supply chain disruptions (10–12).

Traditionally, supply chain security has been established based on trust within the network. This model operates under the premise that entities operating within the perimeter are inherently secure, creating security measures that focus on defending the boundaries of the network. Yet, the rise of sophisticated cyberattacks and insider threats has exposed the limitations of this approach. Adversaries often exploit vulnerabilities in the supply chain ecosystem, potentially compromising critical infrastructure and disrupting the flow of essential medications (8–12).

In reaction to these challenges, Zero trust architecture is becoming more and more popular as an alternative and more proactive approach to cybersecurity. Zero trust, founded on the principle of "never trust, always verify," challenges the traditional perimeter-centric model by supporting continuous verification and authentication of all entities attempting to access resources within the network. This approach mandates strict access controls, least-privilege access, micro-segmentation, and continuous monitoring, irrespective of where the user is located or how trustworthy they think they are. (13,14).

Zero trust architecture has garnered attention across various industries, including information technology, finance, and government, as a means to counter the evolving threat landscape. Applying zero trust principles to supply chain management represents an emerging research frontier with significant implications for enhancing security and resilience. Although the literature on zero trust in supply chain management is still developing, preliminary investigations indicate its potential to address challenges such as insider threats, unauthorized access, and supply chain disruptions (15,16).

In the pharmaceutical sector, where the stakes involve public health and patient safety, applying zero trust principles takes on heightened significance. The industry is marked by intricate multi-tier supply chains involving manufacturers, distributors, wholesalers, pharmacies, and healthcare providers across global networks. Medications' secure and efficient movement through these complex chains is essential to ensure timely and reliable access to vital treatments (17).

Research at the intersection of zero trust and pharmaceutical supply chain management is beginning to uncover insights into the possibilities and challenges of implementing a zero trust approach. Early explorations have demonstrated how zero trust can enhance transparency and accountability by providing granular control over data access and transaction flows. Moreover, it has the potential to facilitate rapid incident response and containment and limit the impact of security breaches and disruptions.

In the forthcoming sections of this paper, We are going to explore the fundamental principles that underpin zero trust architecture, examining how these principles can be tailored and applied to the unique dynamics of the pharmaceutical supply chain. By drawing on existing literature and case studies, We seek to provide a comprehensive understanding of the potential benefits, challenges, and strategies involved in implementing zero trust to fortify the security and resilience of the pharmaceutical supply chain.

**Challenges in the Pharmaceutical Supply Chain**

The pharmaceutical supply chain, a critical lifeline for delivering medications and medical products to patients worldwide, operates within a complex web of interconnected entities, processes, and technologies. While it is indispensable in maintaining public health, it is not immune to various challenges threatening its security, integrity, and resilience. In recent years, the digitalization of supply chain operations, combined with the ever-evolving landscape of cybersecurity threats, has amplified these challenges, requiring a comprehensive assessment of vulnerabilities and the implementation of effective countermeasures (18).

### a) Data Breaches and Cyberattacks

The digitization of the pharmaceutical supply chain has caused a sharp rise in the quantity of sensitive and confidential data being exchanged between various stakeholders. This influx of data presents a tempting target for cybercriminals seeking to take advantage of network weaknesses. Breaches can compromise proprietary research data, patient information, and intellectual property, potentially causing monetary losses, regulatory non-compliance, and reputational damage (19).

### b) Counterfeit Medications and Fraud

The pharmaceutical supply chain's global reach makes it susceptible to the infiltration of counterfeit medications and fraudulent products. Criminal organizations exploit vulnerabilities in the supply chain to introduce substandard or counterfeit drugs, putting patients' health and safety at risk. The lack of transparency and traceability within the supply chain can hinder efforts to identify and mitigate these threats (20).

### c) Supply Chain Disruptions

Disruptions to the pharmaceutical supply chain can arise from various sources, including natural disasters, geopolitical instability, regulatory changes, and labor strikes. The public's health and patient care may be impacted by shortages of necessary pharmaceuticals and medical supplies brought on by these disruptions. The supply chain is interconnected, which increases the impact of disruptions and necessitates the need for resilience and backup plans (21).

### d) Complexity and Lack of Transparency

The intricate nature of the pharmaceutical supply chain, characterized by various tiers of suppliers, distributors, and manufacturers, can result in a lack of visibility and transparency. This opacity can hinder effective tracking and monitoring of product movements, making it difficult to pinpoint issues, enforce quality control, and respond swiftly to emerging threats (22).

### e) Legacy Systems and Interoperability

The pharmaceutical industry often relies on legacy systems and disparate technologies that may not be designed with modern cybersecurity standards in mind. The challenge of integrating these systems with new technologies and platforms can create vulnerabilities that adversaries can exploit.

### f) Regulatory Compliance

The pharmaceutical industry To guarantee the safety, effectiveness, and quality of medications, it functions within a strict regulatory framework. Complying with these regulatory standards while protecting the supply chain from cyber threats presents a considerable challenge. Failure to adhere to regulations may lead to legal repercussions and adversely affect the reputation of the industry and its stakeholders.

Addressing these challenges requires a multifaceted approach that goes beyond traditional security measures. Implementing a zero trust architecture offers a promising avenue for fortifying the pharmaceutical supply chain against these threats. Zero trust can provide a robust framework for enhancing security, transparency, and resilience throughout the supply chain by redefining the principles of trust and focusing on continuous verification and strict access controls (**Figure 1**).

In the subsequent sections of this paper, we will investigate the principles of zero trust architecture and explore how its application can effectively mitigate the challenges outlined above. By aligning zero trust principles with the unique dynamics of the pharmaceutical supply chain, we aim to provide actionable insights and strategies for boosting the resilience and security of this critical industry (18).

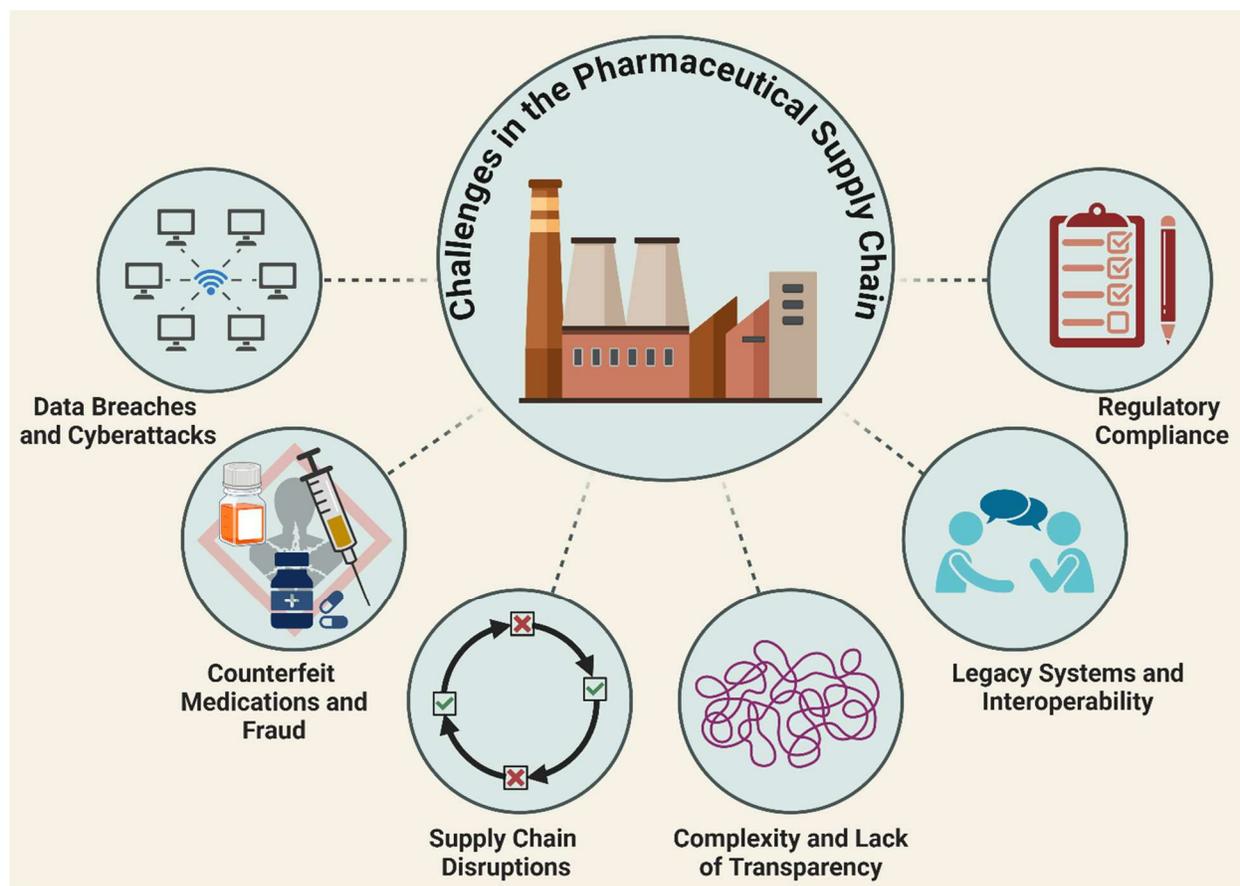

*Figure 1.* Overview of the challenges facing the pharmaceutical supply chain. This figure highlights key issues including data breaches and cyberattacks, counterfeit medications and fraud, supply chain disruptions, complexity and lack of transparency, legacy systems and interoperability, and regulatory compliance. Addressing these challenges requires a comprehensive strategy, with zero trust architecture emerging as a potential solution to enhance security, transparency, and resilience within the supply chain. Created with BioRender.com

**Zero Trust Principles and Framework**

Zero trust architecture questions the fundamental presumption of trust within the network, marking a radical break from conventional perimeter-based security models. Rooted in the philosophy of "never trust, always verify," zero trust instructs continuous verification and authentication of all entities attempting to access network resources, regardless of their location or previous access privileges. By adopting this approach, organizations establish a comprehensive security framework that mitigates the attack surface and minimizes the possible repercussions of breaches.

The core principles that underpin zero trust architecture encompass a series of strategic guidelines and practices that collectively redefine how security is approached within an organization (23):

**Least-Privilege Access:** By enforcing the least-privilege access principle, zero trust makes sure that devices and users are only given the bare minimum of access required to complete their tasks. This prevents the overexposure of sensitive resources and reduces the potential for unauthorized actions (24).

**Micro-Segmentation:** The network is divided into smaller segments, with distinct security controls applied to each segment. This approach restricts lateral movement within the network, preventing attackers from easily moving laterally from one segment to another (25).

**Continuous Authentication and Authorization:** Every user, device, and transaction is continuously authenticated and authorized, even after initial access is granted. This ongoing verification ensures that access remains appropriate and secure throughout the user's session (26).

**Data-Centric Security:** Zero trust focuses on protecting data rather than network perimeters. Data is encrypted, and access controls are applied directly to it, ensuring its confidentiality and integrity regardless of location (27).

**Visibility and Monitoring:** Comprehensive monitoring and logging are integral to zero trust. All network activities are closely monitored, and anomalous behavior is quickly identified and addressed (28).

**Automation and Orchestration:** In response quickly to security incidents and threats, automation is essential to zero trust architecture. Automated responses can isolate compromised systems, mitigate risks, and reduce manual intervention (29).

**Continuous Assessment and Improvement:** Zero trust is a continuous process that calls for continuous assessment, adaptation, and improvement. Regular security assessments and audits ensure that the architecture remains effective and up to date (30).

Implementing zero trust within the pharmaceutical supply chain necessitates a structured framework that aligns with the industry's unique requirements and challenges. The following components constitute a foundational framework for the application of zero trust principles in this context:

**Identifying Critical Assets:** Begin by identifying and classifying critical assets within the pharmaceutical supply chain, including sensitive data, intellectual property, and key operational systems.

**Mapping Data Flows:** Understand how data flows through the supply chain, identifying touchpoints and potential vulnerabilities where zero trust controls can be applied (31).

**Segmentation Strategy:** Develop a segmentation strategy that divides the supply chain into logical segments, each with security controls and access policies (32).

**Authentication and Authorization:** Establish stringent authorization guidelines and robust authentication procedures, confirming users and devices prior to granting access to resources. (33).

**Continuous Monitoring:** Deploy robust monitoring and detection mechanisms to track activities, detect anomalies, and respond to real-time security incidents (34).

**Incident Response Plan:** Formulate a comprehensive incident response plan that explains how to mitigate the impact of security breaches on the supply chain. (35).

**Vendor and Partner Considerations:** Extend zero trust principles to include third-party vendors and partners, ensuring their access aligns with your security policies.

The application of zero trust principles and framework to the pharmaceutical supply chain presents a transformative opportunity to enhance security, resilience, and patient safety. By adopting a zero trust architecture that addresses the unique issues of the industry; stakeholders can create a more secure and agile supply chain ecosystem capable of withstanding evolving cyber threats.

In the subsequent sections, we will explore real-world case studies and practical implementation strategies, showcasing how zero trust principles can be effectively integrated into the pharmaceutical supply chain to mitigate challenges and strengthen overall security.

**Benefits of Zero Trust in the Pharmaceutical Supply Chain**

Adopting zero trust architecture within the pharmaceutical supply chain promises a range of substantial benefits, each contributing to enhancing security, resilience, and overall operational efficiency. By embracing the principles of continuous verification, least-privilege access, and data-centric security, stakeholders in the pharmaceutical industry can fortify their supply chain against existing and emerging cybersecurity threats. The following are key advantages associated with the implementation of zero trust within this critical sector (36):

**Heightened Security and Risk Mitigation**: Zero trust's proactive approach to security significantly reduces the attack surface by requiring continuous verification and strict access controls. This approach mitigates the possibility of unauthorized access, insider threats, and other cyberattacks that can compromise sensitive data, disrupt operations, and compromise patient safety (23,37,38).

**Protection of Sensitive Data**: The data-centric security model of zero trust ensures that sensitive information, including patient data and intellectual property, is safeguarded at all times. By applying granular access controls and encryption directly to the data, the potential for data breaches and unauthorized access is drastically minimized (39,40).

**Resilience and Contingency**: Zero trust's emphasis on micro-segmentation and continuous monitoring enhances the supply chain's resilience against disruptions. In the face of unforeseen events, such as natural disasters or geopolitical instability, rapidly identifying and containing threats ensures minimal impact on the flow of medications and medical supplies (41,42).

**Transparency and Traceability**: Implementing zero trust enables robust visibility into data and transaction flows throughout the supply chain. This transparency facilitates comprehensive tracking and auditing, quickly identifying anomalies, unauthorized access, or suspicious activities (43,44).

**Adaptability to Emerging Threats**: Zero trust architecture's dynamic nature enables companies to adjust to emerging threats quickly. Automated responses and real-time monitoring empower stakeholders to recognize and react to emerging vulnerabilities, mitigating the potential for damage and disruption (45).

**Regulatory Compliance and Auditing**: Zero trust's rigorous access controls and continuous monitoring align with regulatory requirements within the pharmaceutical industry. By enforcing strict data protection measures, Organizations are able to show with assurance that they are adhering to rules and industry standards. (46,47).

**Streamlined Incident Response**: The automation and orchestration inherent in zero trust architecture streamline incident response processes. Risks and the amount of time needed to handle security incidents, downtime, and associated costs are decreased by automated threat detection and response systems.

**Cost Savings**: While the initial implementation of zero trust may require an investment in technology and resources, the long-term benefits include potential cost savings through improved security, reduced risk of breaches, and streamlined operations (48).

**Trustworthy Partner Relationships**: As the pharmaceutical supply chain often involves collaboration with external partners and vendors, implementing zero trust principles enhances trust and confidence in these relationships. It ensures that each entity accessing the supply chain ecosystem adheres to the same rigorous security standards (49).

**Future-Proofing the Supply Chain**: By embracing a zero trust framework, the pharmaceutical industry positions itself to address emerging security challenges and advances in cyber threats. This future-proofing enables the industry to adapt to evolving risks and technologies proactively.

Incorporating zero trust principles into the pharmaceutical supply chain architecture represents a significant step toward fortifying the industry's security posture. Applying these principles aligns with the industry's critical mission of ensuring patient safety and public health, safeguarding sensitive data, and maintaining the uninterrupted flow of medications to those who depend on them (**Figure 2**) (50,51).

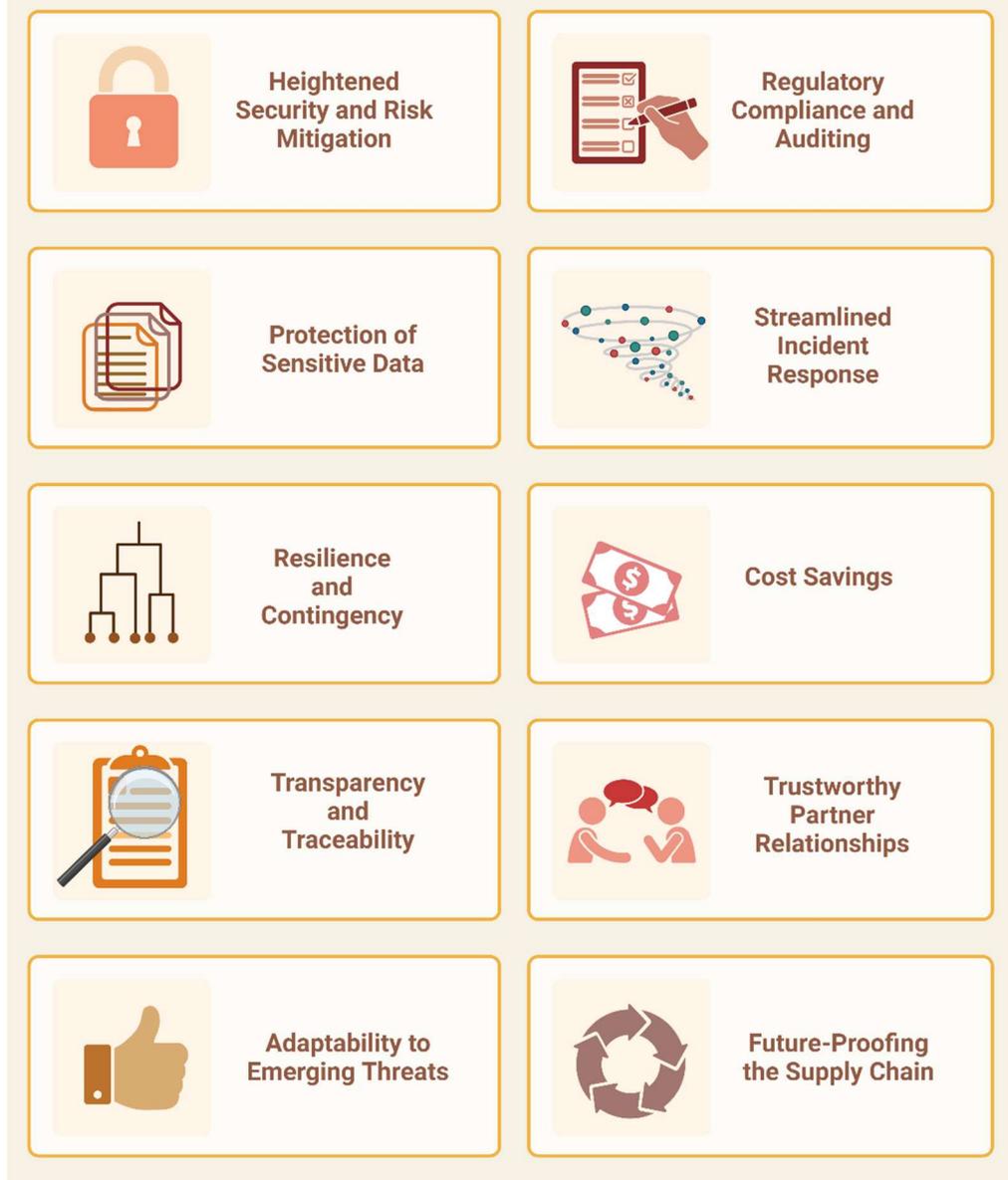

***Figure 2.*** *Illustration of the benefits of implementing zero trust architecture within the pharmaceutical supply chain. The figure highlights key advantages, including heightened security and risk mitigation, protection of sensitive data, enhanced resilience, transparency and traceability, adaptability to emerging threats, and improved regulatory compliance. These benefits collectively contribute to safeguarding patient safety, ensuring the integrity of the supply chain, and positioning the pharmaceutical industry to proactively address future challenges. Created with BioRender.com*

**Five Classes of Drugs**

Depending on the approved medical use of the drug as well as its potential for abuse or dependence, drugs, substances, and some of the compounds used to make them are categorized into five different schedules or categories. When it comes to scheduling a substance, the rate of abuse plays a crucial role. For instance, drugs classified as having high abuse potential and the potential for severe psychological and/or physical dependence are associated with Schedule I. As the drug schedule changes, so does the abuse potential. Schedule V pharmaceuticals reflect the slightest possibility for abuse. One of the most crucial areas within the pharmaceutical industry and its associated supply chains where zero trust can be effectively implemented is in the management of narcotics, high-health-risk drugs, and abusable substances.

**Schedule I**
Substances, drugs, or chemicals classified as Schedule I have a high risk for abuse and no recognized medical use. Heroin, lysergic acid diethylamide (LSD), marijuana (cannabis), methaqualone, peyote, 3,4-methylenedioxymethamphetamine (ecstasy), and cannabis are a few examples of Schedule I drugs (52,53).

**Schedule II**
Drugs classified as Schedule II substances, chemicals, or drugs with a high risk for abuse are those that have the potential to cause severe physical or psychological dependence upon usage. These medications are also thought to be harmful. Medicines classified as Schedule II include combination products (like Vicodin, which contains less than 15 milligrams of hydrocodone per dosage unit), methadone, cocaine, methamphetamine, hydromorphone (like Dilapidid), meperidine (like Demerol), oxycodone (like OxyContin), fentanyl, Dexedrine, Adderall, and Ritalin (52,53).

**Schedule III**
Substances, drugs, or chemicals classified as Schedule III have a moderate to low potential for causing physical or psychological dependence. Drugs on Schedule III have a higher potential for abuse than those on Schedule IV but less than those on Schedule I and II. Products like Tylenol with codeine, ketamine, anabolic steroids, and testosterone that have less than 90 milligrams of codeine per dosage unit are examples of Schedule III drugs (52,53).

**Schedule IV**
Substances, drugs, or chemicals classified as Schedule IV have a minimal likelihood of dependence and abuse potential. Xanax, Soma, Darvon, Darvocet, Valium, Ativan, Talwin, Ambien, and Tramadol are a few drugs that are listed as Schedule IV (52,53).

**Schedule V**
Drugs, substances, or chemicals listed in Schedule V are those that have a lower potential for abuse than those listed in Schedule IV and are prepared foods that contain trace amounts of specific narcotics. Drugs listed in Schedule V are typically used as analgesics, antitussives, and antidiarrheal medications. Cough preparations containing less than 200 mg of codeine per 100 milliliters (Robitussin AC), Lomotil, Motofen, Lyrica, and Parepectolin are a few examples of Schedule V medications (52,53).

Ensuring drug safety is a paramount priority as it has a direct impact on public health. Both researchers and industrialists widely recognize that a reliable drug tracing system is crucial for ensuring drug safety throughout the entire pharmaceutical production process. Medication traceability systems offer several advantages, including enhancing patient safety by safeguarding against counterfeit drugs and decreasing operational expenses and time (**Figure 3**) (54).

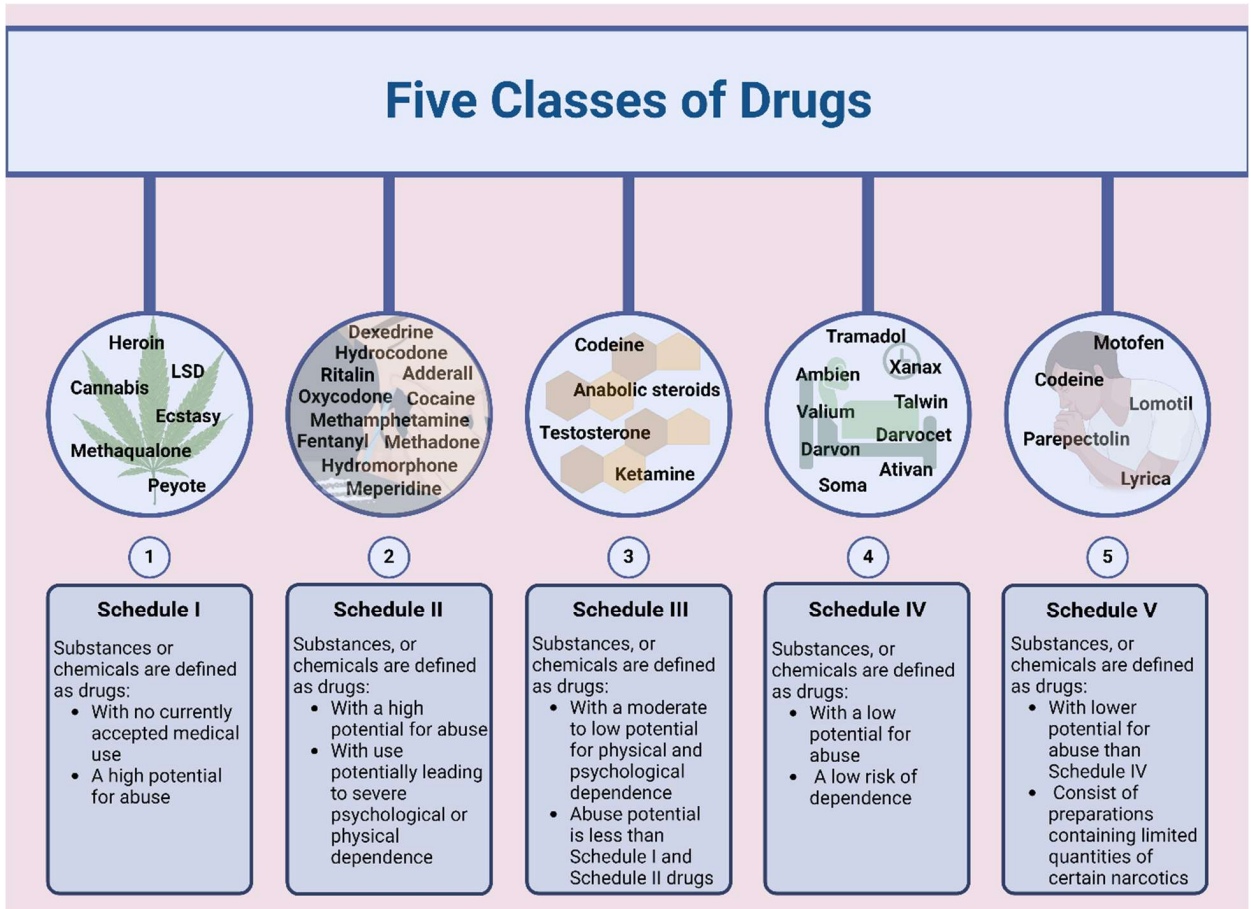

*Figure 3.* Classification of drugs into five schedules based on their medical use and potential for abuse. Schedule I drugs have the highest potential for abuse and no accepted medical use, while Schedule V drugs have the lowest potential for abuse and are often used in medical treatments. The figure illustrates the decreasing potential for abuse and dependence from Schedule I to Schedule V, highlighting the importance of stringent management practices, such as zero trust, in the pharmaceutical industry to ensure the safety and integrity of high-risk substances. Created with BioRender.com

**Future Directions and Research Opportunities**

The pharmaceutical industry is undergoing continuous evolution within the framework of digital transformation and cybersecurity challenges; the integration of zero trust architecture provides a forward-looking approach that holds the potential to shape the future of supply chain security and resilience. This section explores several prospective directions and research possibilities that can further advance the understanding and implementation of zero trust within the pharmaceutical supply chain.

1. **Advanced Threat Intelligence and Analytics**: Future research can delve into developing advanced threat intelligence systems and analytical tools tailored to the pharmaceutical supply chain. These systems can improve the identification of new threats and offer predictive insights that direct preventive security actions by utilizing AI and machine learning. (3,55,56).

2. **Blockchain and Distributed Ledger Technology**: Exploring the synergies between zero trust architecture and emerging technologies like blockchain and distributed ledger can yield innovative solutions for enhancing transparency, traceability, and data integrity within the pharmaceutical supply chain. Research can investigate how these technologies can complement and reinforce zero trust principles (57,58).

3. **Supply Chain Resilience Metrics**: Developing quantitative metrics to assess the resilience of the pharmaceutical supply chain under a zero trust framework is a fertile area for research. By quantifying the impact of security enhancements, organizations can better understand the return on investment and make informed decisions about resource allocation (17,59).

4. **Human-Centric Security Considerations**: While zero trust emphasizes technology and automation, research can delve into the human factors aspect of implementing and maintaining a zero trust architecture. Understanding the roles, responsibilities, and training requirements for personnel within the pharmaceutical supply chain is crucial for successful adoption (3,28).

5. **Interoperability and Standardization**: Investigating interoperability challenges and potential standards for implementing zero trust across diverse pharmaceutical supply chain stakeholders can streamline adoption and collaboration. Research can explore best practices for ensuring seamless integration of zero trust principles across different organizations and systems (60,61).

6. **Economic and Societal Impacts**: Further research can assess the broader economic and societal impacts of zero trust implementation in the pharmaceutical supply chain. This includes evaluating the potential reduction in costs associated with security breaches, as well as the societal benefits of ensuring consistent access to critical medications (62).

7. **Regulatory Considerations**: As zero trust implementation may intersect with regulatory frameworks specific to the pharmaceutical industry, research can explore aligning zero trust principles with evolving regulations and standards, ensuring compliance without compromising security (46,63).

8. **Dynamic Threat Modeling**: Developing dynamic threat modeling methodologies specific to the pharmaceutical supply chain can aid in identifying vulnerabilities, predicting potential attack vectors, and informing proactive mitigation strategies (64,65).

9. **Collaboration and Information Sharing**: It is essential to look into methods for safe cooperation and data exchange within the ecosystem of the pharmaceutical supply chain. Research can focus on establishing trusted communication channels and data exchange protocols that align with zero trust principles (66–68).
10. **Case Studies and Best Practices**: In-depth case studies detailing the successful implementation of zero trust within specific segments of the pharmaceutical supply chain can provide valuable insights and best practices for industry stakeholders considering adoption (23,42,69).

Continued research in these areas will help us comprehend the possibility of zero trust architecture on a deeper level within the pharmaceutical supply chain. By addressing these future directions and research opportunities, the industry can harness the transformative power of zero trust to ensure the pharmaceutical supply chain ecosystem's security, resilience, and reliability in the years to come (70).

**Conclusion and Future Research**

The pharmaceutical supply chain, a cornerstone of public health and patient well-being, stands at the intersection of critical operations and cybersecurity challenges. As digitalization continues to reshape the industry, the imperative to safeguard sensitive data, ensure uninterrupted access to vital medications, and mitigate emerging threats has never been more pressing. For these issues, the adoption of zero trust architecture emerges as a transformative approach that redefines security paradigms and fortifies the foundations of the pharmaceutical supply chain.

This paper has investigated the multifaceted landscape of implementing zero trust architecture within the pharmaceutical supply chain. It has highlighted the vulnerabilities and risks that permeate the industry, from data breaches and counterfeit medications to supply chain disruptions and legacy system vulnerabilities. Through an in-depth examination of zero trust principles and their application, this paper has illustrated how this approach can effectively address these challenges and elevate the security and resilience of the pharmaceutical supply chain ecosystem.

The benefits of zero trust are manifold. By embracing continuous verification, least-privilege access, micro-segmentation, and data-centric security, the pharmaceutical industry can elevate its security posture to withstand the complexities of the modern threat landscape. Improved protection of sensitive data, enhanced resilience against disruptions, and streamlined incident response are just a few of the advantages that zero trust architecture offers.

Implementing zero trust within the pharmaceutical supply chain is an aspiration and an actionable imperative. Case studies from the real world have shown the successful integration of zero trust principles, underscoring its practical applicability and potential for transformative change. As the industry navigates an era of rapid technological advancements and evolving cyber threats, zero trust provides a future-proof framework that can adapt to the shifting landscape and enable continuous innovation.

Numerous examples of this specialized application are evident in the practices of pharmacies and pharmaceutical chains. Before a pharmacy accepts a shipment from a distributor, they implement strict verification protocols. These could include multi-factor authentication (MFA) for the distributor's delivery personnel, blockchain-based verification of the product's origin, and real-time checks on the authenticity of the medicine using serialized tracking. This ensures that the pharmacy is only receiving verified products from trusted and authenticated sources, reducing the risk of counterfeit medicines entering the supply chain. The pharmacy restricts access to its inventory management system to only those employees who need it for their role. Distributors and manufacturers are granted limited access only to the specific data required to fulfill their part of the process, such as shipment schedules or product quantities. By minimizing access, the pharmacy reduces the potential attack surface, preventing unauthorized personnel from tampering with inventory records or shipment details. The pharmacy implements continuous monitoring and anomaly detection across its supply chain network. They use AI-based analytics to detect unusual patterns, such as unexpected shipment delays or changes in delivery routes. If any anomalies are detected, the system immediately triggers an investigation or halts the distribution process. This approach assumes that a breach could happen at any time and ensures that the pharmacy can quickly respond to and mitigate any potential threats before they impact the supply chain.

In conclusion, the incorporation of zero trust architecture within the pharmaceutical supply chain signifies a dedication to patient safety, data security, and the integrity of critical medical operations. By adopting this paradigm, the industry takes a pivotal step towards ensuring the reliability of the supply chain, safeguarding public health, and advancing the broader goals of healthcare delivery. As the pharmaceutical sector embraces the principles of zero trust, it charts a course toward a more secure, resilient, and agile future, where the trust that matters most is that which is continuously verified and earned.

**Acknowledgment**

We would like to thank the Anchor Research Team (ART) for their support.

**Conflicts of interest**

The authors declare that they have no competing interests.

**Author contribution**

S.G., G.S., B.G. and M.L. performed the research. S.G. and G.S. designed the research study. M.R.T contributed in designing illustrations. S.G., G.S. and B.G. analysed the data. B.G. and M.L. wrote the paper.

**Funding**
This research received no specific grant from funding agencies in the public, commercial, or not-for-profit sectors.

# References

1. Shah N. Pharmaceutical supply chains: key issues and strategies for optimisation. Comput Chem Eng. 2004 Jun;28(6–7):929–41.

2. Uchenna Joseph Umoga, Enoch Oluwademilade Sodiya, Olukunle Oladipupo Amoo, Akoh Atadoga. A critical review of emerging cybersecurity threats in financial technologies. International Journal of Science and Research Archive. 2024 Feb 28;11(1):1810–7.

3. Syed NF, Shah SW, Shaghaghi A, Anwar A, Baig Z, Doss R. Zero Trust Architecture (ZTA): A Comprehensive Survey. IEEE Access. 2022;10:57143–79.

4. Szymanski TH. The "Cyber Security via Determinism" Paradigm for a Quantum Safe Zero Trust Deterministic Internet of Things (IoT). IEEE Access. 2022;10:45893–930.

5. Asad M, Otoum S. Integrative Federated Learning and Zero-Trust Approach for Secure Wireless Communications. IEEE Wirel Commun. 2024 Apr;31(2):14–20.

6. Kumar S, Mallipeddi RR. Impact of cybersecurity on operations and supply chain management: Emerging trends and future research directions. Prod Oper Manag. 2022 Dec 1;31(12):4488–500.

7. Wright J. Healthcare cybersecurity and cybercrime supply chain risk management. 2023;

8. Mackey TK, Nayyar G. Digital danger: a review of the global public health, patient safety and cybersecurity threats posed by illicit online pharmacies. Br Med Bull. 2016;118(1):110–26.

9. Möller DPF. Cybersecurity in digital transformation. In: Guide to Cybersecurity in Digital Transformation: Trends, Methods, Technologies, Applications and Best Practices. Springer; 2023. p. 1–70.

10. Hassija V, Chamola V, Gupta V, Jain S, Guizani N. A survey on supply chain security: Application areas, security threats, and solution architectures. IEEE Internet Things J. 2020;8(8):6222–46.

11. Solfa FDG. Impacts of cyber security and supply chain risk on digital operations: evidence from the pharmaceutical industry. International Journal of Technology, Innovation and Management (IJTIM). 2022;2(2):18–32.

12. Urciuoli L, Männistö T, Hintsa J, Khan T. Supply chain cyber security–potential threats. Information & Security: An International Journal. 2013;29(1).

13. Zanasi C, Russo S, Colajanni M. Flexible zero trust architecture for the cybersecurity of industrial IoT infrastructures. Ad Hoc Networks. 2024;103414.

14. Azad MA, Abdullah S, Arshad J, Lallie H, Ahmed YH. Verify and trust: A multidimensional survey of zero-trust security in the age of IoT. Internet of Things. 2024;101227.

15. Federici F, Martintoni D, Senni V. A zero-trust architecture for remote access in industrial IoT infrastructures. Electronics (Basel). 2023;12(3):566.

16. do Amaral TMS, Gondim JJC. Integrating Zero Trust in the cyber supply chain security. In: 2021 Workshop on Communication Networks and Power Systems (WCNPS). IEEE; 2021. p. 1–6.

17. Collier ZA, Sarkis J. The zero trust supply chain: Managing supply chain risk in the absence of trust. Int J Prod Res. 2021;59(11):3430–45.

18. Nguyen A, Lamouri S, Pellerin R, Tamayo S, Lekens B. Data analytics in pharmaceutical supply chains: state of the art, opportunities, and challenges. Int J Prod Res. 2022;60(22):6888–907.

19. Solfa FDG. Impacts of Cyber Security and Supply Chain Risk on Digital Operations: Evidence from the Pharmaceutical Industry. International Journal of Technology, Innovation and Management (IJTIM). 2022 Oct 27;2(2).

20. Mackey TK, Liang BA, York P, Kubic T. Counterfeit drug penetration into global legitimate medicine supply chains: a global assessment. Am J Trop Med Hyg. 2015;92(Suppl 6):59.

21. Tang CS. Robust strategies for mitigating supply chain disruptions. International Journal of Logistics: Research and Applications. 2006;9(1):33–45.

22. Montecchi M, Plangger K, West DC. Supply chain transparency: A bibliometric review and research agenda. Int J Prod Econ. 2021;238:108152.

23. Stafford VA. Zero trust architecture. NIST special publication. 2020;800:207.

24. Brazhuk A, Fernandez EB. An Abstract Security Pattern for Zero Trust Access Control. In: Proceedings of the 29th Conference on Pattern Languages of Programs. 2022. p. 1–5.

25. Basta N, Ikram M, Kaafar MA, Walker A. Towards a zero-trust micro-segmentation network security strategy: an evaluation framework. In: NOMS 2022-2022 IEEE/IFIP Network Operations and Management Symposium. IEEE; 2022. p. 1–7.

26. da Silva GR, Macedo DF, dos Santos AL. Zero trust access control with context-aware and behavior-based continuous authentication for smart homes. In: Anais do XXI Simpósio Brasileiro em Segurança da Informação e de Sistemas Computacionais. SBC; 2021. p. 43–56.

27. Scarfone K, Souppaya M. Data Classification Practices: Facilitating Data-Centric Security Management. 2021;

28. Sarkar S, Choudhary G, Shandilya SK, Hussain A, Kim H. Security of zero trust networks in cloud computing: A comparative review. Sustainability. 2022;14(18):11213.

29. Cao Y, Pokhrel SR, Zhu Y, Doss R, Li G. Automation and orchestration of zero trust architecture: Potential solutions and challenges. Machine Intelligence Research. 2024;1–24.

30. Yeoh W, Liu M, Shore M, Jiang F. Zero trust cybersecurity: Critical success factors and A maturity assessment framework. Comput Secur. 2023;133:103412.

31. Hong S, Xu L, Huang J, Li H, Hu H, Gu G. SysFlow: Toward a Programmable Zero Trust Framework for System Security. IEEE Transactions on Information Forensics and Security. 2023;18:2794–809.

32. Simpson WR, Foltz KE. Network segmentation and zero trust architectures. In: Lecture Notes in Engineering and Computer Science, Proceedings of the World Congress on Engineering (WCE). 2021. p. 201–6.

33. Yao Q, Wang Q, Zhang X, Fei J. Dynamic access control and authorization system based on zero-trust architecture. In: Proceedings of the 2020 1st international conference on control, robotics and intelligent system. 2020. p. 123–7.

34. Meng L, Huang D, An J, Zhou X, Lin F. A continuous authentication protocol without trust authority for zero trust architecture. China Communications. 2022;19(8):198–213.

35. Tsai M, Lee S, Shieh SW. Strategy for Implementing of Zero Trust Architecture. IEEE Trans Reliab. 2024;

36. Kayhan H. Ensuring Trust In Pharmaceutical Supply Chains By Data Protection Through A Design Approach To Blockchains.

37. Ghasemshirazi S, Shirvani G, Alipour MA. Zero Trust: Applications, Challenges, and Opportunities. arXiv preprint arXiv:230903582. 2023;

38. Dhiman P, Saini N, Gulzar Y, Turaev S, Kaur A, Nisa KU, et al. A Review and Comparative Analysis of Relevant Approaches of Zero Trust Network Model. Sensors. 2024;24(4):1328.

39. Ahmed I, Nahar T, Urmi SS, Taher KA. Protection of sensitive data in zero trust model. In: Proceedings of the international conference on computing advancements. 2020. p. 1–5.

40. Greenwood D. Applying the principles of zero-trust architecture to protect sensitive and critical data. Network Security. 2021;2021(6):7–9.

41. Hurley J. Zero-trust is not enough: Mitigating data repository breaches. In: Proceedings of the ICCWS 2023 18th International Conference on Cyber Warfare and Security. 2023.

42. Khan MJ. Zero trust architecture: Redefining network security paradigms in the digital age. World Journal of Advanced Research and Reviews. 2023;19(3):105–16.

43. Buck C, Olenberger C, Schweizer A, Völter F, Eymann T. Never trust, always verify: A multivocal literature review on current knowledge and research gaps of zero-trust. Comput Secur. 2021;110:102436.

44. Muhammad T, Munir MT, Munir MZ, Zafar MW. Integrative Cybersecurity: Merging Zero Trust, Layered Defense, and Global Standards for a Resilient Digital Future. International Journal of Computer Science and Technology. 2022;6(4):99–135.

45. Khan MJ. Zero trust architecture: Redefining network security paradigms in the digital age. World Journal of Advanced Research and Reviews. 2023;19(3):105–16.

46. Bobbert Y, Timmermans T. Zero Trust and Compliance with Industry Frameworks and Regulations: A Structured Zero Trust Approach to Improve Cybersecurity and Reduce the Compliance Burden. In: Future of Information and Communication Conference. Springer; 2024. p. 650–67.

47. Migeon JH, Bobbert Y. Leveraging zero trust security strategy to facilitate compliance to data protection regulations. In: Science and Information Conference. Springer; 2022. p. 847–63.

48. Kindervag J. Build security into your network's dna: The zero trust network architecture. Forrester Research Inc. 2010;27:1–16.

49. Tidjon LN, Khomh F. Never trust, always verify: a roadmap for trustworthy ai? arXiv preprint arXiv:220611981. 2022;

50. Collier ZA, Sarkis J. The zero trust supply chain: Managing supply chain risk in the absence of trust. Int J Prod Res. 2021;59(11):3430–45.

51. Li S, Iqbal M, Saxena N. Future industry internet of things with zero-trust security. Information Systems Frontiers. 2022;1–14.

52. Lopez MJ, Preuss C V, Tadi P. Drug enforcement administration drug scheduling. In: StatPearls [Internet]. StatPearls Publishing; 2023.

53. Miles RJ. Drug Scheduling Paradoxes. Volume II. 2019;57.

54. Liu X, Barenji AV, Li Z, Montreuil B, Huang GQ. Blockchain-based smart tracking and tracing platform for drug supply chain. Comput Ind Eng. 2021;161:107669.

55. Zhang J, Zheng J, Zhang Z, Chen T, Tan Y an, Zhang Q, et al. ATT&CK-based Advanced Persistent Threat attacks risk propagation assessment model for zero trust networks. Computer Networks. 2024;245:110376.

56. Karabacak B, Whittaker T. Zero Trust and Advanced Persistent Threats: Who Will Win the War? In: International Conference on Cyber Warfare and Security. 2022. p. 92–101.

57. Liu Y, Hao X, Ren W, Xiong R, Zhu T, Choo KKR, et al. A blockchain-based decentralized, fair and authenticated information sharing scheme in zero trust internet-of-things. IEEE Transactions on Computers. 2022;72(2):501–12.

58. Dhar S, Bose I. Securing IoT devices using zero trust and blockchain. Journal of Organizational Computing and Electronic Commerce. 2021;31(1):18–34.

59. Ismail S, Moudoud H, Dawoud D, Reza H. Blockchain-Based Zero Trust Supply Chain Security Integrated with Deep Reinforcement Learning. 2024;

60. Li S, Iqbal M, Saxena N. Future industry internet of things with zero-trust security. Information Systems Frontiers. 2022;1–14.

61. Ferretti L, Magnanini F, Andreolini M, Colajanni M. Survivable zero trust for cloud computing environments. Comput Secur. 2021;110:102419.

62. Ray PP. Web3: A comprehensive review on background, technologies, applications, zero-trust architectures, challenges and future directions. Internet of Things and Cyber-Physical Systems. 2023;

63. Bobbert Y, Scheerder J. Zero trust validation: from practical approaches to theory. Sci J Res Rev. 2020;2(5):830–48.

64. Yao Q, Wang Q, Zhang X, Fei J. Dynamic access control and authorization system based on zero-trust architecture. In: Proceedings of the 2020 1st international conference on control, robotics and intelligent system. 2020. p. 123–7.

65. Jin Q, Wang L. Zero-trust based distributed collaborative dynamic access control scheme with deep multi-agent reinforcement learning. EAI Endorsed Transactions on Security and Safety. 2020;8(27).

66. N'goran KR, Brou APB, Pandry KG, Tetchueng JL, Kermarrec Y, Asseu O. Zero Trust Security Strategy for Collaboration Systems. In: 2023 International Symposium on Networks, Computers and Communications (ISNCC). IEEE; 2023. p. 1–6.

67. Liu Y, Hao X, Ren W, Xiong R, Zhu T, Choo KKR, et al. A blockchain-based decentralized, fair and authenticated information sharing scheme in zero trust internet-of-things. IEEE Transactions on Computers. 2022;72(2):501–12.

68. Zhang H, Zhang Z, Chen L. Toward zero trust in 5G industrial internet collaboration systems. Digital Communications and Networks. 2024;

69. Papakonstantinou N, Van Bossuyt DL, Linnosmaa J, Hale B, O'Halloran B. A zero trust hybrid security and safety risk analysis method. J Comput Inf Sci Eng. 2021;21(5):050907.

70. Ma Z, Chen X, Sun T, Wang X, Wu YC, Zhou M. Blockchain-Based Zero-Trust Supply Chain Security Integrated with Deep Reinforcement Learning for Inventory Optimization. Future Internet. 2024;16(5):163.